\documentclass[fleqn,11pt]{wlscirep}
\usepackage{multirow}
\usepackage{wrapfig}
\usepackage{float}
\usepackage{tabularx}%
\usepackage[title]{appendix}%
\usepackage{booktabs}%
\usepackage[T1]{fontenc}
\usepackage{makecell}%
\usepackage{graphicx}
\usepackage{subfig}
\usepackage{siunitx}
\usepackage{hyperref}
\usepackage{lineno}
\usepackage{changes}

\title{Urban mobility enables deprivation bubble breaking in Indian and Mexican cities}
\author[1,2,3,*]{Yuan Liao}
\author[2]{Federico Delussu}
\author[2]{Sílvia de Sojo}
\author[2]{Laura Alessandretti}
\author[2,4,5]{Antonio Desiderio}
\affil[1]{Department of Space, Earth and Environment, Chalmers University of Technology, Gothenburg, Sweden}
\affil[2]{Department of Applied Mathematics and Computer Science, Technical University of Denmark, Richard Petersens Plads, 2800 Lyngby, Denmark}
\affil[3]{Department of Human Geography, Lund University, Lund, Sweden}
\affil[4]{Centro Ricerche Enrico Fermi, Via Panisperna 89a, 00184 Rome, Italy}
\affil[5]{ISI Foundation, Via Chisola 5, 10126 Turin, Italy.}
\affil[$*$]{Corresponding author: \href{mailto:yuan.liao@keg.lu.se}{yuan.liao@keg.lu.se}.}
\begin{abstract}

Urban deprivation is traditionally measured using static, residence-based indicators, capturing the socioeconomic, demographic, and spatial conditions of neighborhoods.
However, this approach overlooks how daily movement allows residents to navigate the city, potentially exposing them to opportunities that differ significantly from their residential environments.
To bridge this gap, we quantify the extent of bubble breaking -- travel to less deprived areas -- by analyzing mobile phone mobility networks combined with satellite-derived deprivation indices across 64 cities in India and Mexico.
We find that residents of deprived areas systematically travel to better-off locations to meet daily needs, exhibiting a compensatory mobility pattern that significantly exceeds expectations derived from gravity models based on population and road networks.
This residual bubble breaking (the part gravity models can not explain) is associated with a tension in the built environment: while high local amenity diversity allows residents to satisfy needs locally, high amenity density and positive spillovers from neighboring areas is associated with movement across socioeconomic boundaries.
Overall, residual bubble breaking reflects the extent to which residents rely on cross-neighborhood mobility to overcome local amenity deficits, a dimension of spatial inequality that residence-based measures leave unobserved.
\end{abstract}
\usepackage{url}
\begin{document}

\flushbottom
\maketitle
\thispagestyle{empty}

%\linenumbers
\section*{Introduction}

Urban deprivation tends to concentrate into spatial and socioeconomic ``bubbles'' -- geographically bounded neighbourhoods with persistently low access to economic and social opportunities~\cite{un2004challenge,kuffer2016slums}.
This pattern is particularly pronounced in low- and middle-income countries, where poverty, limited access to services, discrimination, and institutional barriers give rise to spatially and socially segregated pockets such as slums and informal settlements~\cite{bettencourt_growth_2007,sugar_urban_2021,mcgranahan_inclusive_2016
,sahasranaman_urban_2019}.
These ``bubbles'' are commonly understood as \emph{spatial deprivation traps}: daily life is assumed to be confined to the immediate area, with residents relying on locally based informal work and facing mobility constraints~\cite{geurs_accessibility_2004}.
These constraints are thought to limit access to broader employment opportunities~\cite{moretti_real_2013}, public services~\cite{larrabee_sonderlund_racialized_2022}, and social infrastructure~\cite{fraser_great_2024}. \par

A common interpretation of such deprivation traps is that many such neighbourhoods accumulate structural disadvantages that are difficult to escape.
Yet, this interpretation rests almost entirely on static, residence-based data~\cite{tiznado2023unequal}. 
Most deprivation research relies on satellite images and other residence-based indicators that show where disadvantage is located but not how residents move through the city in their daily lives~\cite{luo2022urban}.
The characteristics of deprived urban areas have been widely studied in relation to hazard exposure \cite{fox2024integrating}, health outcomes \cite{headen2018associations}, and educational attainment \cite{lund2020moving}.
However, most deprivation research does not account for how individuals experience different levels of deprivation as they move beyond their residential neighbourhoods~\cite{luo2022urban}. \par

In contrast, the field of urban segregation and inequalities has increasingly moved beyond residence-based measures toward mobility-based approaches to understand the socioeconomic mixing individuals \textit{experience} across the spaces they visit~\cite{nilforoshan2023human,renninger2025us,liao2025socio}. 
This shift was enabled by the availability of mobile phone data that captures individuals' movements, with most empirical evidence to date drawn from high-income countries~\cite{liao2025socio}.
Yet deprivation research rarely incorporates mobility data, leaving us without evidence on activity-based deprivation exposure --- the levels of deprivation individuals encounter across their daily destinations.
Classical theories of urban economics~\cite{barthelemy2026mathematical, alonso_location_1964} and accessibility~\cite{hansen1959accessibility,kain1968housing, glaeser_why_2008} suggest that jobs and services are concentrated in more advantaged areas, generating structural pressures on residents of deprived neighbourhoods to travel beyond their immediate surroundings to meet their daily activity needs.
This motivates the hypothesis that daily mobility systematically shifts individuals' exposure away from residential deprivation and toward more advantaged urban environments.
Testing this hypothesis at scale is especially urgent in low- and middle-income countries, where inequality and deprivation are salient issues affecting large populations~\cite{ureta2008move,bautista2020commuting}, but where country-level mobility data have historically been unavailable. \par

Here, we address this gap by testing whether, in highly unequal urban contexts, individuals tend to systematically travel to areas less deprived than their residential neighbourhoods --- a pattern we describe using the term ‘bubble breaking’, drawing on prior uses of the phrase in the segregation literature~\cite{park2021we}.
We do so using a mobile-phone dataset from 64 cities in India and Mexico, two countries that differ markedly in urban form, infrastructure, and patterns of deprivation.
We find consistent evidence that individuals tend to travel to less deprived areas, and that this tendency is stronger than expected under a gravity-based mobility model~\cite{zipf1946p1p2} that accounts only for population distribution and the road network ~\cite{carpio2021multimodal,aiello2025urban}.
Additionally, we show that three key features of the urban environment---overall urban form~\cite{schwarz2010urban,burton_measuring_2002,barrington-leigh_global_2020}, local neighbourhood opportunities~\cite{moro_mobility_2021,yabe2023behavioral,yue_measurements_2017}, and mobility-related constraints~\cite{weiss_global_2018,giacomin_road_2015,salazar_miranda_desirable_2021}---help explain variation in the extent of bubble breaking across neighbourhoods and cities.

\section*{Results}
To test the bubble-breaking hypothesis at scale, we considered mobility networks capturing travel in India and Mexico. 
Network nodes represent geographical areas of approximately 5 $km^2$ defined using the H3 discrete global grid system~\cite{uber2024h3github} at resolution 7 (see Methods). 
A directed link between two nodes exists if at least one trip occurs between the corresponding areas, and link weights correspond to the total number of trips generated by individuals on that connection over the period from November 1, 2019, to March 1, 2020. 
These flows are derived from aggregated smartphone GPS traces provided by Cuebiq as part of the NetMob 2024 Data Challenge \cite{milusheva2024netmob24} (see ``Mobility Network'' in the Methods).
We identified 64 spatially disjoint clusters of at least 25 contiguous areas where mobility data were available (see ``City Boundary'' in the Methods for details). \par

Then, we assign multiple attributes to each area to characterize its social and spatial context. 
First, we quantify socioeconomic deprivation using NASA's Global Gridded Relative Deprivation Index (GRDI)~\cite{CIESIN_2022}. 
The GRDI ranges from 0 (least deprived) to 100 (most deprived), based on a combination of sociodemographic and satellite-derived indicators, including child dependency ratios, infant mortality rates, subnational human development index, building footprints, and nighttime lights (see ``Deprivation'' in the Methods section, Fig.~\ref{fig:res1a}a, b, and Supplementary Fig.~\ref{fig:a1}).
This indicator is globally defined and applied in absolute terms, without normalization to country- or city-specific deprivation distributions; this choice is reflected in the higher overall deprivation levels observed in New Delhi compared to Greater Mexico City (Fig.~\ref{fig:res1a}a, b).
Additionally, we describe each area using population density from the Global Human Settlement Layer (GHSL)~\cite{schiavina_ghs-pop_2023} and the number and diversity of local amenities derived from Point of Interest (POI) data from Overture Maps~\cite{overturemaps2024} (see the Methods section for more details).

\subsection*{Individuals break residential deprivation bubbles}
\begin{figure}[!ht] %[H]
    \centering
    \includegraphics[width=1\linewidth]{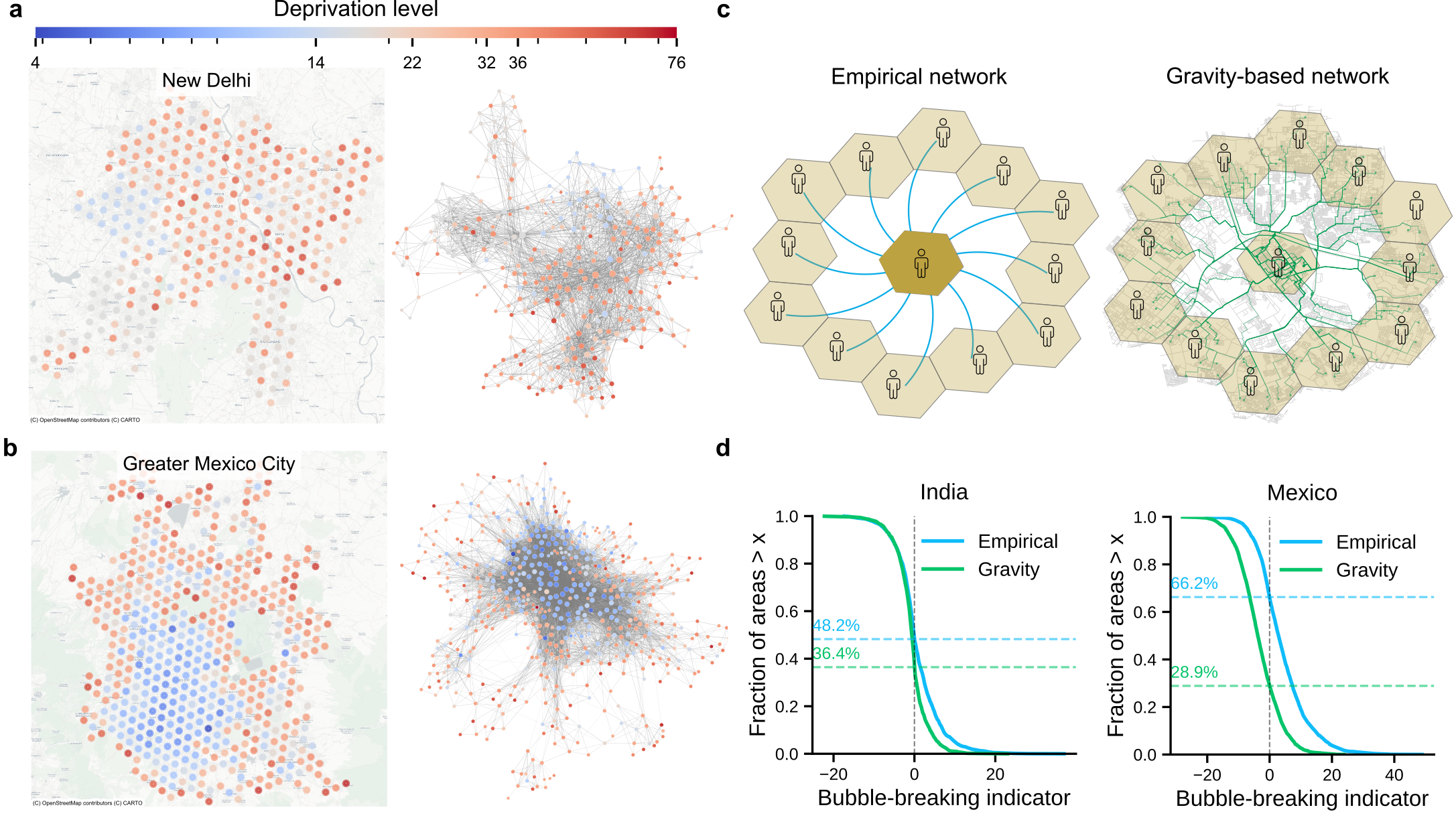}
    \caption{\textbf{Bubble breaking patterns via the lens of mobility network.} 
\textbf{a}-\textbf{b}, Spatial distribution of residential deprivation in New Delhi and Greater Mexico City. 
Each dot represents the centroid of an area (hexagonal cell) and is colored by its deprivation level (blue = low, red = high), using a common color scale across cities. 
The right panel shows empirical mobility networks constructed from observed inter-area trips, where nodes correspond to areas and edges represent aggregated travel flows between them. 
\textbf{c}, Schematic comparison of deprivation exposure under two mobility networks: the empirical network (left) and a gravity-based model constrained by road-network travel costs and population distribution (right). 
\textbf{d}, Cumulative (complementary) distribution of the bubble-breaking indicator across areas for the empirical data (blue) and the gravity-model baseline (green), where the fraction represents the share of areas with indicator values greater than $x$. The dashed vertical line marks the neutral threshold (0), while horizontal dashed lines indicate the share of areas with positive values, i.e., areas that are exposed to less deprived areas than their residential level.
}
    \label{fig:res1a}
\end{figure}

We quantify the extent to which individuals break out of their residential deprivation ``bubbles'' by comparing the deprivation levels of the areas they visited with those of their home areas. 
Specifically, we define a \textit{Bubble Breaking Indicator} as $D_i = R_i - E_i$. 
Here, $E_i$ denotes the deprivation exposure of area $i$, calculated as the weighted deprivation level of the areas connected to $i$ by outgoing links in the mobility network, where weights correspond to the volume of travel along each link.
$R_i$ represents the residential deprivation level of area~$i$ (see ``Bubble Breaking Indicator'' in the Methods section).
A bubble-breaking indicator $D_i$ above 0 indicates that the population living in area $i$ is exposed to areas less deprived than their residential environment, reflecting a break from their residential bubble. 
Conversely, a value below 0 suggests exposure to more deprived areas compared to where they reside.
To assess the extent to which bubble-breaking exceeds expectations based solely on population distribution and travel-time constraints, we compare empirical flows with those predicted by a standard gravity model of mobility~\cite{zipf1946p1p2}, in which travel between two areas increases with their populations and decreases with the travel time between them along the road network (see Methods for more details on the gravity model formulation).
Specifically, for each area $i$, we compute the empirical bubble-breaking ($D_i^e$) and its gravity-model expectation ($D_i^t$), using the observed and the gravity-expected mobility networks, respectively (Fig.~\ref{fig:res1a}a-b).
They reflect the extent to which individuals break out of their residential deprivation bubble (Fig.~\ref{fig:res1a}c). 
In constructing gravity-model expectation ($D_i^t$), we applied different travel-time constraints (10–45 minutes; 1–2 hours), reflecting the potential exposure implied by a gravity-based mobility network.
Throughout this study, we report results using a 45-minute travel-time threshold for gravity-based deprivation exposure (see Supplementary Fig. \ref{fig:threshold}), as the residual bubble breaking stabilizes beyond this threshold. \par

Focusing first on the empirical values $D_i^e$, we observe pronounced empirical bubble breaking in both India and Mexico (Fig.~\ref{fig:res1a}d, blue markers). 
Overall, 78\% of areas across the studied cities exhibit positive bubble breaking, and the distribution of $D_i^e$ lies significantly above zero in both countries (India: p = 0.002; Mexico: p < 0.001; Wilcoxon signed-rank test).
In contrast, under the gravity-model baseline, the expected bubble-breaking index $D_i^t$ across areas 
is not statistically different from zero (p=1; Fig.~\ref{fig:res1a}d, green markers).
Together, these results demonstrate that daily mobility systematically exposes residents to environments that are less deprived than their home areas, beyond what would be expected from population size and travel-time constraints alone. \par

\subsection*{Bubble breaking is more pronounced in high deprivation contexts}

\begin{figure}[!ht] %[H]
    \centering
    \includegraphics[width=1\linewidth]{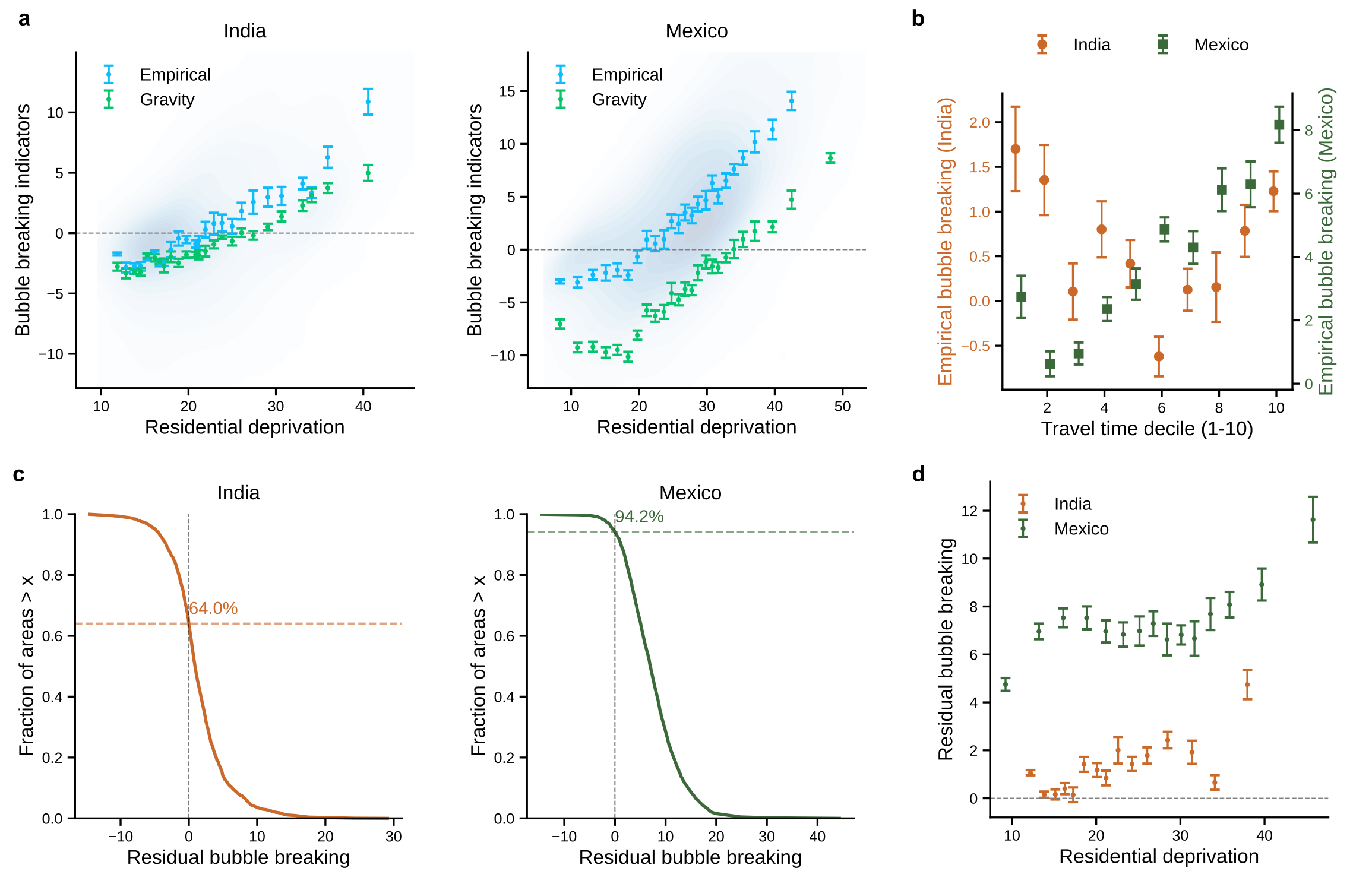}
    \caption{\textbf{Bubble breaking patterns are associated with deprivation level.} 
\textbf{a}, Bubble breaking indicators (D) vs. residential deprivation level (R). 
The blue density plot illustrates the distribution of areas across empirical vs. residential deprivation combinations. 
For the bubble-breaking indicator $D$, $D > 0$ indicates exposure to less deprived areas, while $D < 0$ indicates exposure to more deprived areas. 
\textbf{b}, Empirical bubble breaking vs. travel time duration decile groups. 
\textbf{c}, Cumulative (complementary) distributions of the residual bubble breaking in India (left) and Mexico (right). 
Dashed lines indicate the fraction of areas with positive bubble breaking, i.e., areas that are exposed to a less deprived area than their residential level.
\textbf{d}, Residual bubble breaking vs. deprivation:
Areas are grouped into 15 quantile bins by residential deprivation. 
Error bars indicate standard errors of the median, estimated via bootstrap.
}
    \label{fig:res1b}
\end{figure}

While empirical bubble breaking is positive on average across areas, its direction and magnitude vary systematically with residential deprivation.
We find a positive correlation between the empirical bubble-breaking and residential deprivation (corr. coefficient: 0.45 for India and 0.60 for Mexico, $p < 0.001$, see Fig.~\ref{fig:res1b}a, blue markers).
Additionally, two contrasting regimes emerge across the deprivation spectrum.
Residents of relatively more deprived areas ($R_i > 25$) are exposed, on average, to less deprived environments through their daily mobility ($D_i^e > 0$), whereas residents of less deprived areas ($R_i < 25$) are exposed, on average, to relatively more deprived areas ($D_i^e < 0$). 
When restricting the analysis to low-deprivation neighbourhoods, however, the association between residential deprivation and bubble breaking weakens substantially (corr. coefficient: 0.092 for India and –0.10 for Mexico, $p < 0.05$), suggesting that differences in exposure are lower among the least deprived areas.
Additionally, in Mexico, stronger bubble breaking is associated with longer travel times (Fig.~\ref{fig:res1b}b), indicated by mobility networks' records of median trip duration between areas. \par

Taken together, these results reveal a pronounced asymmetry in deprivation exposure: residents of more deprived areas systematically shift their exposure toward better-off environments through daily mobility, whereas residents of less deprived areas show no consistent tendency to avoid more deprived contexts, potentially reflecting a lower need to travel far to meet day-to-day activity demand. \par

The dependence of bubble breaking on deprivation may stem from fundamental structural features of cities, such as population distribution and the street network, as evidenced by the similar pattern produced by the gravity-based indicator
 (Fig.~\ref{fig:res1b}a).
To isolate deviations from these structural expectations, we introduce the \emph{residual bubble breaking} index, $\text{RBB}_i=D_i^e-D_i^t=E_i^t - E_i^e$ for area $i$ residents (see ``Residual Bubble Breaking'' in the Methods section).
This index captures the extent to which empirical bubble breaking differs from the gravity model's prediction. 
Intuitively, a positive RBB indicates stronger-than-expected bubble breaking, meaning that residents are exposed to less deprivation than population distribution and travel time constraints alone would predict. \par

In both India and Mexico, the residual is significantly positive (Fig.~\ref{fig:res1b}c, $p < 0.001$), and increases with residential deprivation (Fig.~\ref{fig:res1b}d, corr. coefficient: 0.2 for India and 0.16 for Mexico, $p < 0.001$). 
This pattern suggests that mobility from more deprived neighbourhoods disproportionately bridges socioeconomic divides, exceeding what would be predicted by mobility accessibility constraints alone.

\subsection*{Urban features explain bubble breaking through direct and spillover effects}\label{sec:results_spillovers}
Having established that bubble-breaking patterns vary across deprivation levels, we next examine the factors underlying this heterogeneity. 
Specifically, we model the residual bubble-breaking index (RBB) for each area using urban features as explanatory variables. \par

\begin{figure}[!ht] %[H]
    \centering
    \includegraphics[width=1\linewidth]{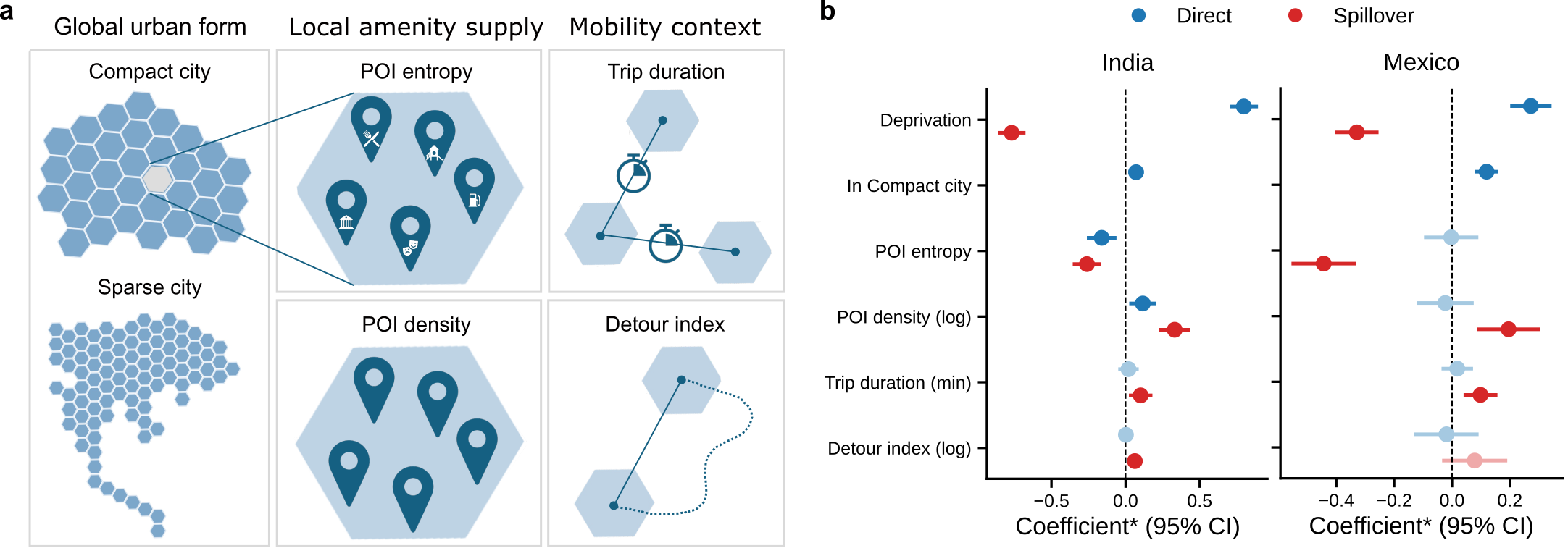}
    \caption{\textbf{Multilevel predictors of residual bubble breaking.} 
\textbf{a}, Residual bubble breaking is affected by three complementary dimensions: global urban form, local amenity supply, and mobility context.
\textbf{b}, SLX regression coefficients (point estimates with 95\% CIs) for India and Mexico, distinguishing \emph{direct} (blue) from \emph{spillover} (red) effects. 
Light colours indicate insignificant coefficients with $p>0.05$.
Direct effects quantify how local conditions predict an area's residual bubble breaking; spillover effects quantify how neighbouring characteristics predict the same outcome.
    }
    \label{fig:res31}
\end{figure}

Besides residential deprivation, we consider three key features: global urban form, local amenity supply, and transport networks (Fig.~\ref{fig:res31}a).
Global urban form is captured by city \emph{compactness}~\cite{schwarz2010urban}, distinguishing contiguous (\textit{compact}) from sprawling (\textit{sparse}) cities (Supplementary Fig.~\ref{fig:urban_form}), with more compact forms linked with short trips.
Local amenity supply is described by the entropy of POI categories and their density. 
The former is interpreted as capturing functional mix, which relates to capacity of satisfying daily needs locally~\cite{omwamba2025assessment}, while the latter may capture activity intensity and spatial concentration.
Mobility context is quantified through median travel time between areas and the detour index.
Lower travel times and detours indicate more efficient connectivity, reducing the friction associated with crossing deprivation boundaries~\cite{salazar_miranda_desirable_2021}.
Further details on variable definitions are provided in ``Explaining residual bubble breaking'' in the Methods, with descriptive statistics reported in Supplementary Table~\ref{tab:stats}. \par

Urban environments are spatially interdependent, so it is essential to account for the influence that surrounding areas exert on any given area~\cite{sampson2019neighbourhood}.
To explicitly capture these interdependences, we employ a Spatially Lagged X (SLX) model (see ``Modeling direct and spillover effects on residual bubble breaking'' in the Methods section), which incorporates spatial lags of the explanatory variables to quantify spillover effects from neighbouring areas~\cite{halleck2015slx}.
In this framework, a positive coefficient indicates that higher values of the explanatory variable are associated with stronger-than-expected bubble breaking---namely, lower empirical deprivation exposure relative to the gravity baseline---whereas a negative coefficient indicates weaker-than-expected bubble breaking. 
Full model results are reported in Supplementary Table~\ref{tab:slx_comparison}.\par

While the SLX models achieve moderate explanatory power (with an adjusted $R^2$ of 0.347 for India and 0.145 for Mexico), we find that urban features show robust associations with residual bubble breaking (see Fig.~\ref{fig:res31}b).
Urban compactness exhibits a positive direct association in both countries (India: 0.71, 95\% CI = 0.32–1.10; Mexico: 1.42, 95\% CI = 1.01–1.83), suggesting that more spatially contiguous and better-integrated city structures tend to coincide with stronger-than-expected bubble breaking. 
Compact cities also show stronger area-level activity persistence than sparse cities during COVID-19 (Supplementary Figure \ref{fig_supp_disc}b).
In contrast, transport network features---including median trip duration and the detour index---do not exhibit a statistically significant direct association with residual bubble breaking in either country (Supplementary Table~\ref{tab:slx_comparison}). 
% However, spatial spillovers tell a more nuanced story areas surrounded by areas with higher travel times or greater network inefficiencies tend to exhibit stronger residual bubble breaking. 
This pattern shifts when considering spatial spillovers: areas adjacent to those with longer travel times or greater network inefficiencies exhibit stronger residual bubble breaking.
Despite this, transport network characteristics contribute only modestly to overall model performance (Supplementary Table~\ref{tab:slx_var_contributions}), suggesting that travel friction alone is not systematically associated with deviations from gravity-based exposure. 
% Should we make a mention/relation to the Gravity model (relation to travel time)
The most substantial effects by the variance explained (see Supplementary Table \ref{tab:slx_var_contributions}) are produced by residential deprivation itself and by the amenity supply. 

\subsection*{Deprivation exhibits opposing direct and spillover effects on residual bubble breaking}\label{sec:trapping_effect}
Residential deprivation has opposing direct and spillover effects on bubble breaking (Fig.~\ref{fig:res31}b).
Focusing on the direct effect, higher deprivation is a robust positive predictor of residual bubble breaking in both countries (with regression coefficients in India: 0.44, 95\% CI=0.40 -- 0.49; Mexico: 0.15, 95\% CI=0.11 -- 0.19).
This is consistent with the pattern that residents of more deprived areas tend to be exposed to less deprived areas than the gravity baseline would predict.
In contrast, deprivation in neighbouring areas exerts a strong negative spillover effect (India: -0.52, 95\% CI=-0.57 -- -0.46; Mexico: -0.24, 95\% CI=-0.28 -- -0.19). 
This implies that residual bubble breaking declines in areas surrounded by high deprivation, indicating reduced exposure to less deprived areas than expected given the surrounding spatial context. \par

\begin{figure}[!ht] %[H]
    \centering
    \includegraphics[width=1\linewidth]{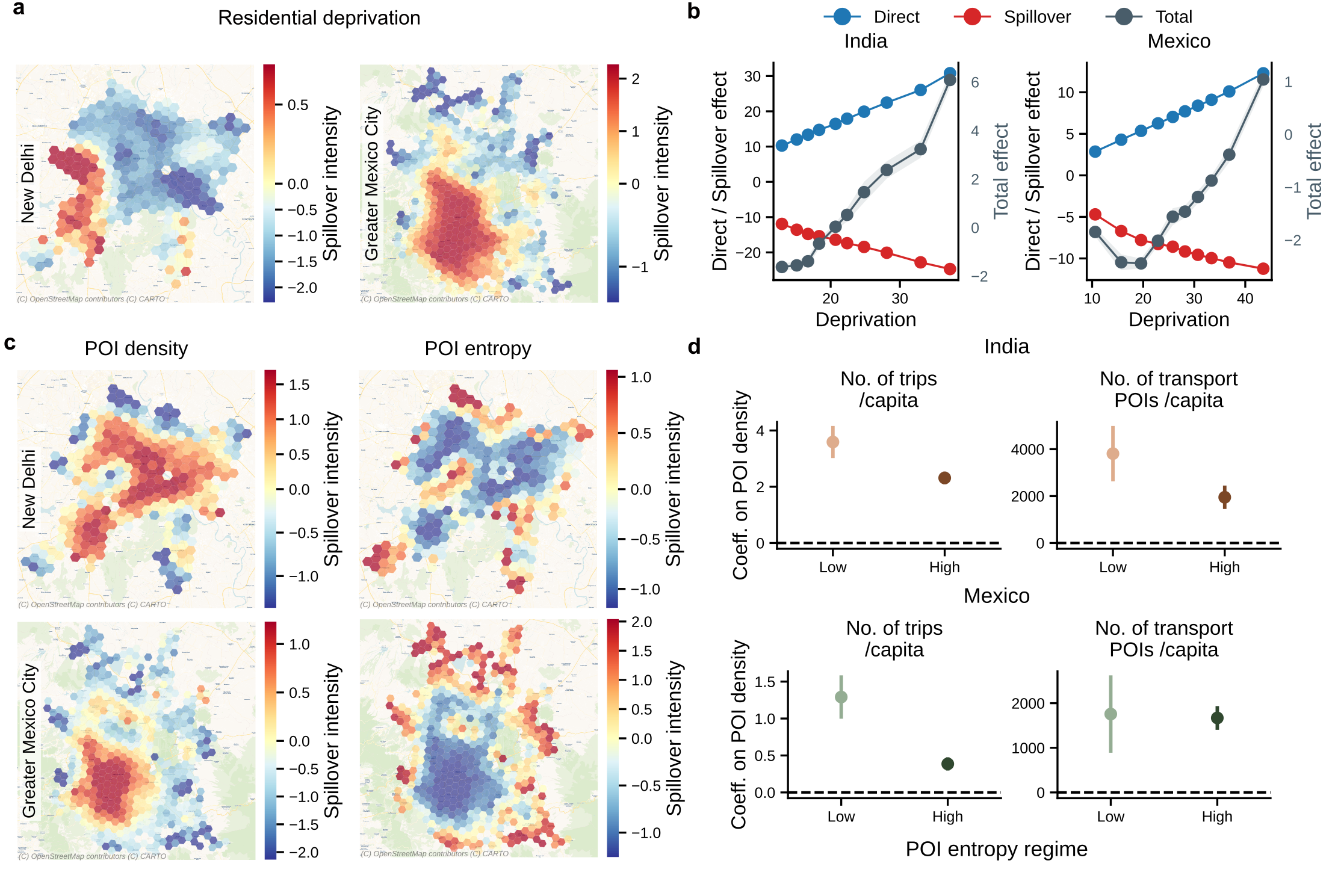}
    \caption{\textbf{The effects of residential deprivation and amenity supply on residual bubble breaking.} 
\textbf{a}, Spatial distribution of spillover intensity (standardized) for residential deprivation in New Delhi and Greater Mexico City. Cooler colours indicate stronger negative spillover effects (``trapping'' in disadvantaged clusters), whereas warmer colours reflect positive spillovers. 
\textbf{b}, Marginal effects of deprivation on residual bubble breaking across ten deciles of residential deprivation. 
Direct (blue), spillover (red) and total sum of both effects (grey).
Solid lines represent median marginal effects; shaded bands indicate 95\% CIs around the medians using bootstrap.
\textbf{c}, Spatial distribution of spillover intensity (standardized) for POI density and diversity in New Delhi and Greater Mexico City. 
\textbf{d}, Contrast in how mobility volume and transport infrastructure explain POI density under low- and high-entropy regimes in India and Mexico. 
Points show OLS coefficients with 95\% confidence intervals.
    }
    \label{fig:res32}
\end{figure}

Together, these opposing effects suggest that while residents of deprived areas may individually exceed gravity-based predictions, spatial clusters of deprivation constrain such deviations, pointing to a potential ``trapping'' dynamic at the level of groups of areas.
These interactions become evident when visualizing spillover intensity across entire cities (see Fig.\ref{fig:res32}).
In New Delhi and Greater Mexico City, clusters of highly deprived areas (red areas in Fig.\ref{fig:res1a}a) correspond to stronger negative spillover effects (blue areas in Fig.\ref{fig:res32}a), whereas some areas adjacent to less deprived regions (blue in Fig.\ref{fig:res1a}a) exhibit stronger positive spillovers (red in Fig.\ref{fig:res32}a).
% Some areas close to the less deprived areas (blue in Fig.\ref{fig:res1a}a) tend to have stronger positive spillover effects (red areas in Fig.\ref{fig:res32}a).
The latter act as ``mobility corridors'' in which residents are more likely to break beyond their residential deprivation context. 
These spatial configurations highlight that residual bubble breaking emerges from interactions across neighbourhood boundaries rather than from local deprivation alone. \par

To assess whether the behavior observed in individual cities generalizes across the country, we examine how direct and spillover effects interact with residential deprivation at the area level.
Figure~\ref{fig:res32}b reveals a divergence between the direct and spillover marginal effects along the deprivation spectrum, consistent with the observed trapping effect.
% Indeed, the trapping effect is apparent in Fig.~\ref{fig:res32}b, which shows a divergence between the direct effect vs. the spillover marginal effect of deprivation along the deprivation spectrum.
As deprivation increases, the direct effect indicates a stronger tendency for individuals to break out of their focal deprivation bubble. 
However, the spillover effect simultaneously becomes negative, indicating that surrounding deprivation is increasingly associated with lower bubble breaking. 
%Federico - interpretation of contrasting results (maybe can be moved to the discussion if it makes sense)
This contrast may partly reflect the imposed trip-duration threshold: limiting trips to under 45 minutes likely confines movements within deprived clusters, strengthening negative spillover effects by reducing the likelihood of crossing deprivation boundaries.
Taken together (see the total marginal effect in Fig.~\ref{fig:res32}b, grey), more deprived contexts are characterized by a stronger pull to break out, coupled with an increasingly powerful trapping effect from neighbouring areas.

\subsection*{Opposing roles of amenity entropy and density}\label{sec:density_diversity}

Amenity-related features --- POI entropy and POI density --- show opposing associations with residual bubble breaking (Fig.\ref{fig:res31}b): diversity is negatively associated, whereas density is positively associated.
This contrast is evident in both direct and through spatial spillovers. 
In India, local POI entropy shows a negative direct association (-0.83, 95\% CI = -1.28 to -0.38), while local POI density has a positive effect (0.47, 95\% CI = 0.14 to 0.75).
Across neighbouring areas, higher POI entropy is linked to lower residual breaking (India: -2.12; Mexico: -3.02), whereas higher POI density is linked to higher residual breaking (India: 1.80; Mexico: 0.99).
When facing disruptions like COVID-19, these functionally diverse neighborhoods, i.e., those with a high POI entropy, are most resilient: they could sustain activity locally (see Supplementary Figure \ref{fig_supp_disc}b).
Overall, higher POI entropy is associated with lower residual bubble breaking, while higher amenity density is associated with stronger-than-gravity bubble breaking. \par

Notably, these opposing effects of POI entropy and density contrast with their positive spatial correlation (Supplementary Fig.~\ref{fig:res2}b). 
We hypothesize that this divergence reflects that density and diversity capture different aspects of the built environment~\cite{geurs_accessibility_2004,krehl_urban_2015,yue_measurements_2017}:
POI entropy reflects the range of activities available locally, whereas density proxies movement-oriented centrality--- areas that concentrate trips and transport infrastructure rather than self-contained activity provision.
To disentangle these effects, we stratify areas by whether their POI entropy lies above or below the median (see Supplementary Note \ref{secb:density_puzzle}), and model POI density using mobility indicators: total trips generated per capita (from mobile phone data) and transport-related POIs per capita (from Overture). 
This allows us to distinguish distinct mechanisms underlying amenity concentration and to clarify why density retains a positive association with residual bubble breaking even when controlling for diversity (Fig.~\ref{fig:res32}d). \par

In low-entropy areas, POI density is more strongly associated with deprivation, mobility throughput, and transport infrastructure intensity than in high-entropy areas. 
The association with transport-related POIs nearly doubles in India (low entropy: $\beta=3809$, 95\% CI=2704--4914 vs. high entropy: $\beta=1949$, 95\% CI=1525--2373) and remains substantial in Mexico (low: $\beta=1757$, 95\% CI=929.0--2585 vs. high: $\beta=1669$, 95\% CI=1443--1896).
This pattern indicates that, in low-diversity contexts, POI density proxies activity hubs or transport-oriented areas rather than functionally complete neighbourhoods, which in turn are associated with cross-boundary movement. \par

Together, these results suggest a potential reconciliation of the apparent contradiction: functional diversity may anchor residents locally by enabling local need satisfaction, whereas amenity density may proxy for activity intensity associated with cross-boundary flows. 

\section*{Discussion}\label{sec12}

Across 64 cities in India and Mexico, we find that residents of deprived neighborhoods systematically travel to less deprived areas in their daily mobility, a pattern we call bubble breaking. 
This tendency is stronger than what would be expected based solely on population distribution and road networks. 
The gap between observed and expected mobility is largest in the most deprived areas, suggesting that residents compensate for inadequate local amenities by traveling farther. 
Neighborhood context matters: clusters of deprived areas suppress this compensatory mobility, while proximity to amenity-rich areas facilitates it. 
Functionally diverse neighborhoods reduce the need for such travel, whereas areas with high amenity density, particularly transit hubs, act as destinations that draw people across deprivation boundaries. \par

% Novelty of residual bubble breaking
Residual bubble breaking differs from existing metrics in the types of bubbles it captures. 
Accessibility measures ask what residents can reach but not where they actually go; experienced segregation indices track who people encounter but require individual demographic data rarely available in Global South cities; activity space measures describe how far people travel but ignore the character of destinations~\cite{liao2025socio}. 
We define residual bubble breaking to ask a different question: does observed mobility expose residents to more or less deprivation than distance and population distribution alone would predict? 
This isolates the behavioral residual --- destination choices shaped by employment, social ties, and local amenity supply --- from physical travel constraints, using only aggregate flows and area-level deprivation data. \par

% How these findings relate to prior literature
This study contributes to urban deprivation research by analyzing mobility networks from large-scale mobile phone data in two developing countries.
We find a strong positive link between residential deprivation and bubble breaking, especially in areas with deprivation scores above 25, consistent with prior work showing that disadvantaged populations compensate for inadequate local resources through mobility \cite{kain1968housing}.
In contrast, residents of less-deprived areas show no clear exposure patterns, in line with evidence that adequate local amenities reduce the need for cross-boundary travel \cite{yu2024travel}.
Two main findings help explain these patterns.
First, spillover coefficients often exceed direct coefficients in magnitude, suggesting that bubble breaking is shaped by surrounding conditions: residents in deprived neighborhoods face both a strong pull to escape local disadvantage and neighboring conditions that limit their ability to do so, creating poverty traps where entire clusters remain isolated \cite{sampson2019neighbourhood}.
Second, proximity to better-served areas creates mobility corridors that enable residents to reach opportunities despite moderate local deprivation, suggesting that deprivation persists not only through residential sorting but also through fragmented mobility networks \cite{liao2025socio}.
The contrasting effects of POI diversity (negative) and density (positive) point to a further tension: diverse neighborhoods anchor residents by meeting multiple needs locally \cite{grant2002mixed}, while high amenity density draws people across deprivation boundaries.
Our stratified analysis clarifies this by showing that high-density, low-diversity areas, often transit hubs in more deprived locations, act as mobility generators, implying that how amenities are distributed matters more than their absolute number \cite{yue_measurements_2017}. \par

%\textbf{Limitations and future directions}. 
The present study has a few limitations that motivate further research.
First, our cross-sectional analysis captures bubble-breaking patterns but not causality between the built environment and bubble-breaking behaviors. 
Second, mobile phone data may underrepresent the most vulnerable populations without any devices \cite{wesolowski2013impact}, suggesting the need for further validation.
Third, POI datasets likely overlook informal economy activities that are crucial in developing contexts \cite{luo2022urban, straulino2022uncovering}. 
This poses a limitation, as more deprived areas often host informal economic activities that are underrepresented in POI data. 
As a result, areas that appear to have low functional diversity may in fact sustain vibrant informal economies that remain invisible in our measurements.
Fourth, we cannot observe actual travel modes, a critical issue, as road networks often assume automobile access, while many residents rely on public transit or walking, particularly those with low incomes \cite{liao2025effect,prieto-curiel_abc_2024}. 
Fifth, observed bubble-breaking mobility cannot be interpreted as purely voluntary behavior: the data do not allow us to distinguish whether cross-deprivation travel reflects active choice or is instead compelled by unmet local needs and structural constraints in residents' home neighborhoods.
Finally, the SLX models achieve moderate explanatory power, with unexplained variance likely reflecting unmeasured household-level characteristics, informal economic activities underrepresented in POI data, social network effects, and cross-city heterogeneity in how urban form relates to mobility patterns.
Future research should incorporate longitudinal dynamics, address unmeasured confounders such as income sorting or employment location decisions, disaggregate by trip purpose, validate against ground-truth surveys, and, most critically, assess whether bubble-breaking predicts upward socioeconomic mobility.
\par

% Policy implications. 
Our findings support several urban intervention strategies.
Amenity diversity reduces residual bubble breaking, suggesting that improving local amenity provision in deprived areas, through mixed-use zoning and equitable resource allocation, could reduce compelled travel rather than merely facilitating it \cite{gao2023socio,prieto-curiel_urban_2025}.
However, where local amenity diversity remains low, transport-related POIs play a heightened role in shaping mobility patterns, pointing to the importance of enhancing connectivity where travel remains necessary, for instance by prioritizing affordable public transit over road expansion \cite{pereira2017distributive}.
The prominence of spillover effects also suggests that interventions may be more effective when coordinated across neighboring areas rather than targeting isolated neighborhoods, particularly where deprivation clusters create a trapping effect.
Residual bubble breaking can serve as a diagnostic metric: high values signal spatial mismatches requiring intervention, but also reflect residents' reliance on mobility to compensate for local deficits.
When facing disruptions like COVID-19, areas with high pre-pandemic residual bubble-breaking tend to be more vulnerable to activity collapse, signaling structural dependence on mobility that becomes fragile when movement is restricted, while POI entropy (local functional diversity) emerges as the more robust foundation for resilience (Supplementary Section \ref{secb:covid-19}).
Together, these patterns highlight the interplay between everyday mobility and structural urban conditions, positioning residual bubble breaking as a practical tool for identifying spatial vulnerabilities and guiding equitable urban policy \cite{pandey2022infrastructure}.

\section*{Methods}\label{sec:methods}
\subsection*{Data Description}
Our analysis integrates multiple geospatial and mobility datasets harmonized at the H3 resolution 7 (approximately 5~$\text{km}^2$) \cite{uber2024h3github}. 
Mobility flows are derived from anonymized origin-destination matrices provided by Cuebiq for the NetMob 2024 Data Challenge \cite{zhang2024netmob2024}, capturing weekly travel from November 1, 2019, to December 31, 2020, and strictly contain data from at least 10 unique devices to ensure privacy. 
We defined cities as contiguous spatial components of at least 25 connected areas (Queen contiguity), matching boundaries to GADM administrative regions and WikiData 
\cite{wikidata}.
Socioeconomic context is quantified using NASA's Global Gridded Relative Deprivation Index (GRDI) \cite{CIESIN_2022}, which provides multidimensional deprivation scores (0--100) at a 30 arc-second ($\sim$1~km) resolution derived from six components (child dependency ratio, infant mortality rates, subnational human development index, built-up area, nighttime light intensity, and its temporal slope), complemented by population counts from the Global Human Settlement Layer (GHSL) \cite{schiavina_ghs-pop_2023}. 
Urban features were constructed using Points of Interest (POIs) from Overture Maps \cite{overturemaps2024}---categorized via co-occurrence networks filtered with the disparity filter ($\alpha < 0.05$) \cite{serrano_extracting_2009} and Infomap clustering \cite{rosvall_maps_2008}---and drivable road networks from OpenStreetMap (OSM), restricted to motorways and primary, secondary, and tertiary roads.

\subsection*{Quantifying Residual Bubble Breaking}
To quantify how residents navigate urban socioeconomic divides, we establish empirical and theoretical measures of deprivation exposure. 
We conceptualize residential deprivation exposure as an individual's deprivation bubble, which structurally constrains their ability to move through and engage with other parts of the city and beyond. 
Residential deprivation ($R_i$) is the average GRDI value across 1 km grid centroids intersecting area $i$. Empirical deprivation exposure ($E_i^e$) is the flow-weighted average deprivation of all destinations visited by residents of area $i$, extracted from the observed mobility network.
To assess how well the built environment -- particularly the travel friction imposed by driving road networks -- predicts deprivation exposure, we establish a structural baseline. 
We sample 100 OSM road-network intersections per area, compute average shortest-path travel times $t_{ij}$ between area pairs, and simulate theoretical flows using a gravity model proportional to population products ($P_i$,$P_j$) and travel time. 
Applying a 45-minute threshold, we compute the gravity-based exposure ($E_i^t$), with results stabilizing beyond this threshold (Supplementary Fig.~\ref{fig:threshold}). \par

The \emph{residual bubble breaking} index is then defined as \[ \text{RBB}_i = D_i^e - D_i^t = (R_i - E_i^e) - (R_i - E_i^t) = E_i^t - E_i^e \] capturing the deviation of empirical exposure from built-environment expectations. 
This residual reflects the extent to which empirical bubble-breaking deviates from theoretical expectations, quantifying how actual exposure deviates from the constraints of the physical network. 
Positive values indicate that individuals reach less-deprived areas more than population distribution alone would predict, effectively quantifying how strongly individuals overcome road network barriers to reach areas with varying levels of deprivation.

\subsection*{Explaining Residual Bubble Breaking}
To analyze how local conditions and neighboring environments jointly predict residual bubble breaking, we estimate Spatially Lagged X (SLX) models separately for India and Mexico. 
The SLX framework transparently decomposes direct versus spillover effects without assuming mutually reinforcing outcomes across space, making it robust for regions with disconnected spatial components. 
This decomposition allows us to isolate how the broader spatial fabric --- beyond an area's own characteristics --- contributes to bubble-breaking behaviors within each country. \par

Operating on the H3 grid, we define a row-standardized binary spatial weights matrix $W$ based on first-order Queen contiguity to capture adjacent hexagons. 
This localized construction naturally captures the geometry of each city's built environment while accommodating boundary areas that inherently possess fewer neighbors. 
The model takes the form $\mathbf{y}=\mathbf{X}\beta+W\mathbf{X}\theta+\varepsilon$, where the dependent variable $\mathbf{y}$ is $\text{RBB}_i$. \par

The design matrix $\mathbf{X}$ includes: POI entropy ($H=-\sum_kp_k\log{p_k}$, where $p_k$ is the share of POIs in category $k$); log-transformed POI density (no. of POIs per area); residential deprivation; log-transformed detour index (the ratio of shortest-path network distance to Euclidean distance, averaged over 100 sampled intersection pairs per area); empirical median trip duration (in min, directly provided in the mobility network data); and a binary indicator for city compactness ($4\pi\times\text{Area}/\text{Perimeter}^2$, classified as compact versus sparse relative to the national median). 
Spatially lagged covariates ($W\mathbf{X}$) represent the average characteristics of each area's immediate neighbors. 
Models are estimated using heteroskedasticity-robust OLS with HC3 standard errors to ensure reliable inference under non-Gaussian error structures.

\section*{Declarations}
\textbf{Data Availability}. 
Origin-destination matrices were made available from Cuebiq and the World Bank as part of the NetMob 2024 Data Challenge \cite{zhang2024netmob2024}.
OpenStreetMap data is openly accessible via \url{https://www.openstreetmap.org/}.
NASA’s Global Gridded Relative Deprivation Index (GRDI) was retrieved from \url{https://sedac.ciesin.columbia.edu/data/set/povmap-grdi-v1}.
Global Human Settlement data was retrieved from \url{https://ghsl.jrc.ec.europa.eu/}.
Points of Interest (POIs) were retrieved from Overture Maps Foundation Places dataset, derived from Meta and Microsoft products such as Bing Maps and Facebook pages, accessible via \url{https://overturemaps.org/}. \par

\noindent\textbf{Code Availability}.
The data and code to reproduce the analysis is released on \href{https://github.com/RiegelGestr/ResidualBubbleBreakingNetmob24}{GitHub}.
For inquiries, please contact A.D. \url{antde@dtu.dk}. \par

\noindent\textbf{Acknowledgements}.
The authors would like to acknowledge Cuebiq, the World Bank, and NetMob 2024, through which the data used in this manuscript were made available.
Thanks to the insightful discussions with all the conference participants.
The authors would like to thank Louis Boucherie and Lasse Pelle Skytte Hansen for their contributions to the initial data challenge project, which made this work possible. 
We also gratefully acknowledge Louis Boucherie for his insightful comments and suggestions on an earlier version of this manuscript.

\noindent\textbf{Funding}. Y.L. acknowledges support from the Swedish Research Council (Project Number 2022-06215). Open access funding is provided by Lund University.
A.D. acknowledges support from the Lagrange Project of the ISI Foundation, funded by Fondazione CRT.

\noindent\textbf{Author Contributions}.
F.D. and S.D.S performed the preliminary exploratory analysis.
A.D., L.A. and Y.L. designed the analysis and supervised the project.
A.D. and Y.L. performed the analysis.
A.D., F.D., Y.L., S.D.S made the figures.
All authors discussed the results and contributed to the final manuscript. \par

\noindent\textbf{Competing Interests}.
The authors declare no competing interests.

% \bmhead{Supplementary information}
\renewcommand\refname{References}
\bibliography{bibliography_c}

\newpage
\begin{appendices}

\renewcommand{\thefigure}{\arabic{figure}}
\renewcommand{\thetable}{\arabic{table}}
\renewcommand{\figurename}{Supplementary Figure}
\renewcommand{\tablename}{Supplementary Table}
\setcounter{figure}{0}
\setcounter{table}{0}

\begin{center}
    {\LARGE \bfseries Supplementary Information\\}
    \vspace{1cm}
    {\LARGE \bfseries Urban mobility enables deprivation bubble breaking in Indian and Mexican cities}
\end{center}

\section{Data description}

\begin{figure}[!ht] %[H]
    \centering
    \includegraphics[width=1\linewidth]{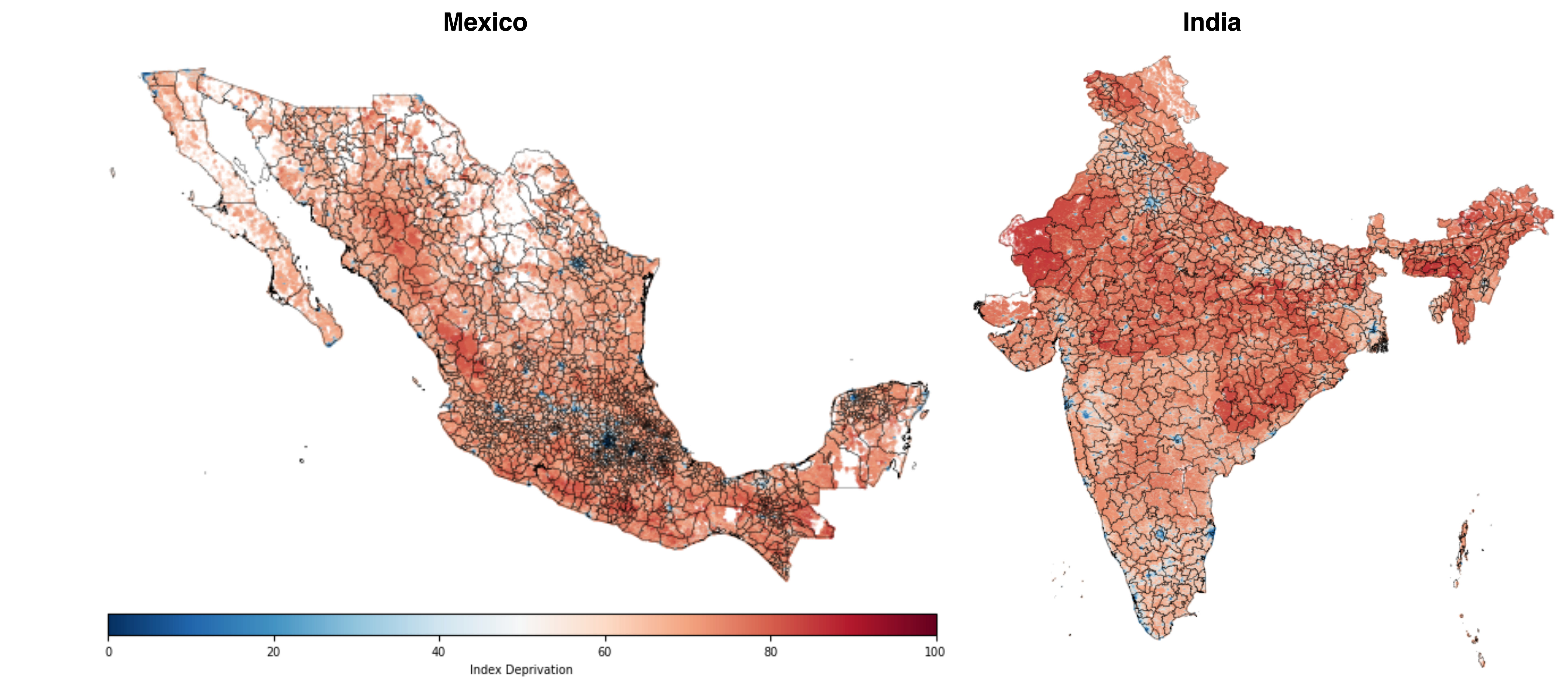}
    \caption{\textbf{Global Gridded Relative Deprivation Index (GRDI). } 
    Spatial distribution of GRDI values for Mexico (left) and India (right). 
    Grid areas have a resolution of 1 km, with color indicating deprivation scores ranging from 0 (lowest) to 100 (highest). Country shapefiles are shown at the GADM level-2 resolution.
    }
    \label{fig:a1}
\end{figure}

Supplementary Fig. \ref{fig:a1} shows the global distribution of the Gridded Relative Deprivation Index (GRDI), highlighting stark spatial inequalities in socioeconomic conditions across regions. 
Higher GRDI values indicate greater levels of relative deprivation at the grid level, allowing for consistent cross-region comparisons. 
The values are derived from satellite and sociodemographic data collected between 2010 and 2020 (see \href{https://www.earthdata.nasa.gov/news/feature-articles/measuring-relative-poverty-deprivation-around-globe}{here} for details on data construction).
This spatialized metric provides the foundational input for our exposure and residual bubble breaking analyses by quantifying deprivation at a granular geographic scale. \par

\section{Time threshold in the gravity model}
To establish a baseline, we applied a gravity model to estimate the expected level of deprivation exposure. 
The model captures the potential exposure individuals would experience given a certain travel-time constraint and population distribution. 
Specifically, we evaluated a series of travel-time thresholds, 10, 15, 20, 30, and 45 minutes, as well as 1 and 2 hours, to estimate the gravity-based deprivation exposure implied by a gravity-based mobility network. \par

\begin{figure}[!ht] %[H]
    \centering
    \includegraphics[width=0.5\linewidth]{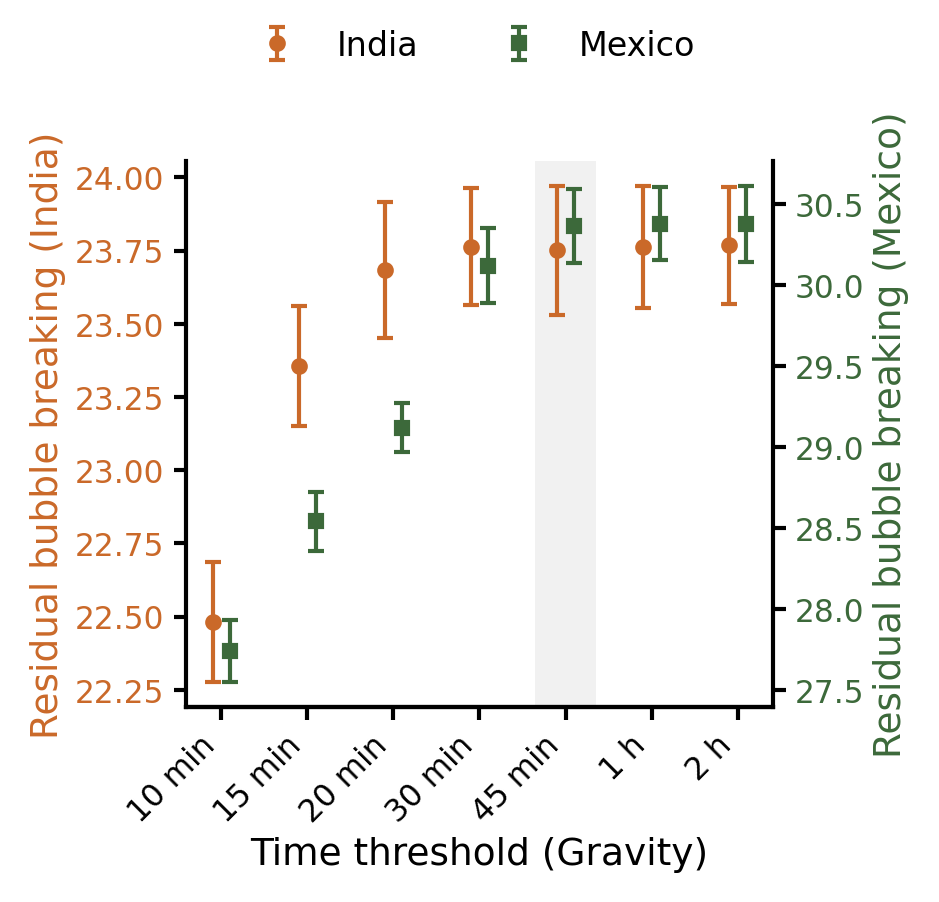}
    \caption{\textbf{Residual bubble breaking vs. travel time threshold in the gravity model.} The main results in the manuscript use the 45-minute threshold, highlighted by the shaded region.}
    \label{fig:threshold}
\end{figure}

Supplementary Fig. \ref{fig:threshold} shows that residual bubble breaking stabilizes once the travel-time threshold in the gravity-based calculation reaches 45 minutes, justifying the selection of the results from a 45-minute time threshold reported in the main manuscript.

\section{Residual bubble breaking links to urban form, amenity, and mobility}\label{sec:corr}

Focusing first on city-level compactness, we compared the distributions of area-level residual bubble-breaking values between compact and sparse cities within each country using the Mann–Whitney-U test. 
Compact cities exhibit significantly higher residual bubble-breaking values than sparse ones in both India ($p < 0.001$) and Mexico ($p < 0.001$). 
Although the effect sizes are modest (Cohen's~$d = 0.26$ for India and~$0.16$ for Mexico), the consistent direction of these differences suggests that more compact urban structures facilitate greater exposure beyond residents' local deprivation contexts. \par

\begin{figure}[!ht] %[H]
    \centering
    \includegraphics[width=1\linewidth]{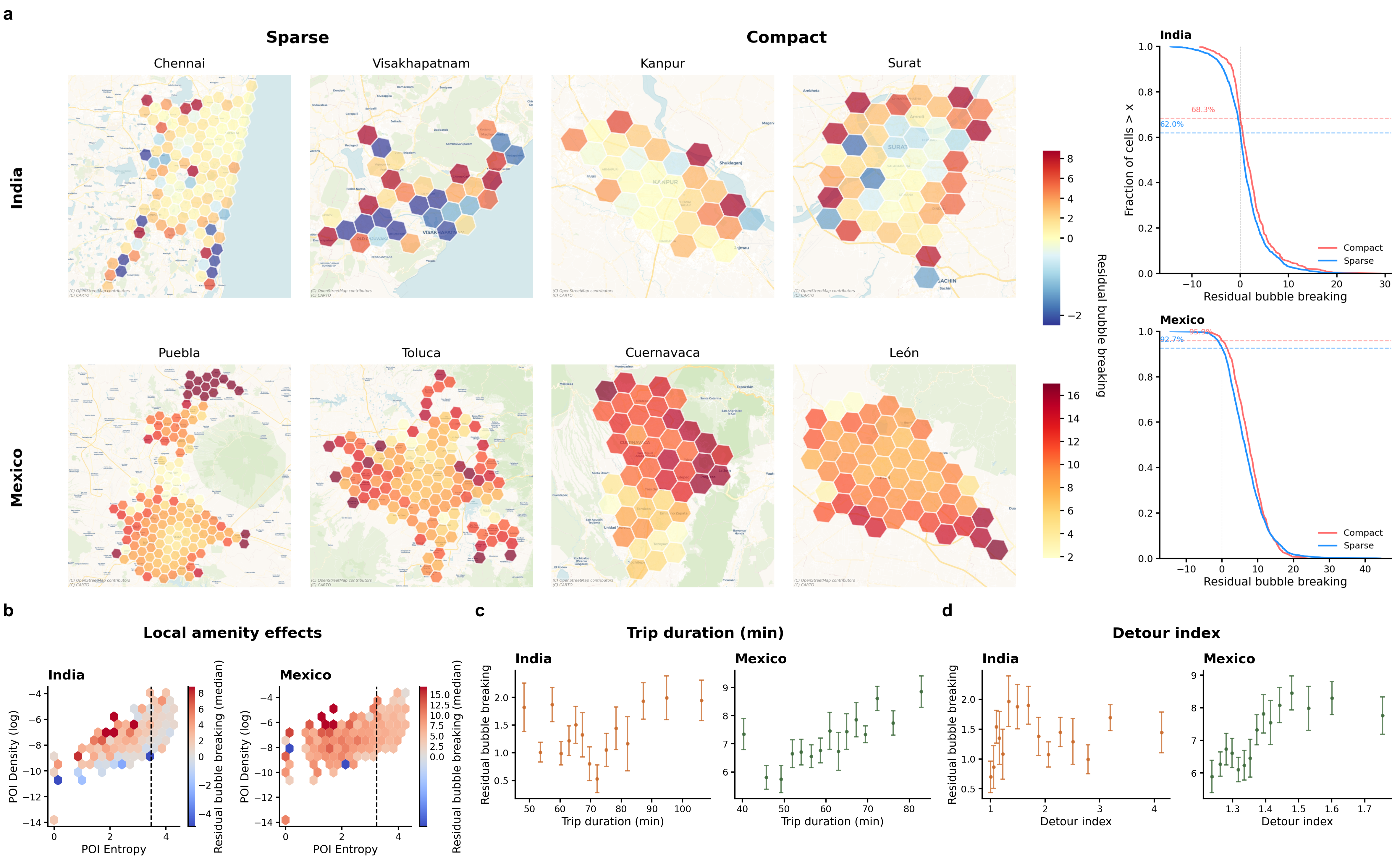}
    \caption{\textbf{Residual bubble breaking and relevant factors.} 
\textbf{a}, Spatial patterns of residual bubble breaking in representative sparse and compact cities in India and Mexico. 
The right panels show the cumulative distribution of residual bubble-breaking values for compact and sparse city types, with dashed lines indicating the shares above zero. 
\textbf{b–d}, Associations between residual bubble breaking and relevant indicators, including \textbf{b} local amenity features: POI entropy and POI density, where vertical lines indicate median values of POI entropy, \textbf{c} average trip duration (min), and \textbf{d} detour index. 
Results are shown separately for India (orange) and Mexico (green). 
Areas grouped into 15 quantile bins by the corresponding factor. 
Error bars indicate median estimation errors.}
    \label{fig:res2}
\end{figure}

Next, we examine area-level patterns across cities by relating residual bubble breaking to local amenity characteristics (Supplementary Fig.~\ref{fig:res2}b), measured by the diversity (entropy) and intensity (density) of points of interest (POIs). 
We observe a general negative correlation between POI density/entropy and residual bubble breaking when examined separately (see Supplementary Fig.~\ref{fig:2a} and Supplementary Table \ref{tab:stats}). 
However, Supplementary Fig.~\ref{fig:res2}b suggests an interaction effect: when POI entropy is below the median level (dashed vertical line), higher POI density is associated with stronger residual bubble breaking in both countries. 
This effect weakens in the higher-entropy regime (> median level). 
These findings suggest non-linear and joint influences of local amenity features, motivating further regression analysis to unpack their combined roles. \par

\begin{figure}[!ht] %[H]
    \centering
    \includegraphics[width=1\linewidth]{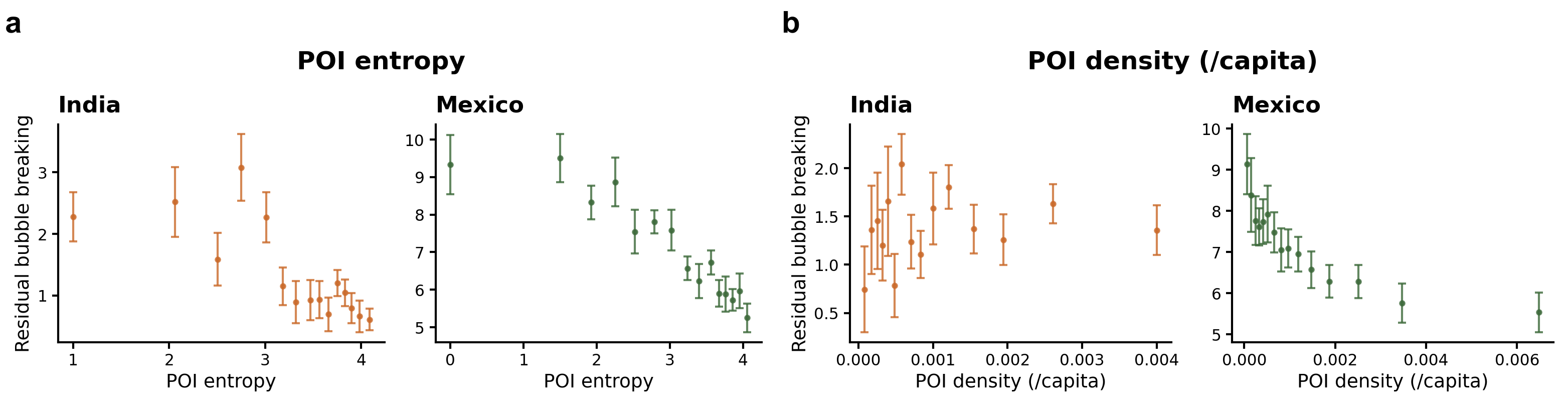}
    \caption{\textbf{Associations between residual bubble breaking and POI features.} \textbf{a} POI entropy. \textbf{b} POI density. 
    Results are shown separately for India (orange) and Mexico (green). 
    Areas grouped into 15 quantile bins by the corresponding factor. 
Error bars indicate median estimation errors.
    }
    \label{fig:2a}
\end{figure}

Beyond urban form and amenity characteristics, mobility patterns also link to residual bubble breaking. 
We examine the role of trip duration, hypothesizing that individuals who travel longer distances are more likely to deviate from the gravity baseline constrained by a 30-minute theoretical travel time. 
Consistent with this expectation, longer trips are associated with greater residual bubble breaking in Mexico, indicating that people become more exposed to less deprived areas than predicted by the gravity model ($\rho = 0.15$, $p < 0.001$; Supplementary Fig.~\ref{fig:res2}c), whereas  
no significant association is observed in India. %, as reflected by a significant negative correlation in Mexico ($\rho = 0.15$, $p < 0.001$) but not in India.
Trip duration, however, captures only how far people travel, not how efficiently they can reach their destinations. 
The other mobility feature considered is the detour index, which quantifies the extent to which actual travel paths along the road network deviate from straight-line distances.
We hypothesize that the detour index may influence residual bubble breaking by constraining individuals' ability to reach less deprived areas. 
The results (Supplementary Fig.~\ref{fig:res2}d) show a weak but significant positive correlation in Mexico ($\rho = 0.14$, $p < 0.001$).
In India, however, the relationship is not statistically significant ($\rho = 0.03$, $p > 0.1$), which may be partly explained by the longer average travel times observed there compared to Mexico---making the detour index a less constraining factor. \par

Overall, the exploratory analysis suggests that global urban form, local amenity supply, and mobility context are associated with residual bubble breaking (Supplementary Table~\ref{tab:stats}), motivating a multivariate spatial modeling approach. \par

\begin{table}[ht]
\centering
\caption{Summary statistics (median and 5--95\% range) and Spearman correlations with residual bubble breaking for explanatory variables in India and Mexico. $^{***}p<0.001$. $^a$Urban compactness measured at the city level, comprising 26 cities in India and 38 cities in Mexico.}
\begin{tabular}{l r r r r}
\toprule
\textbf{Country}/\textbf{Variable} & \textbf{Median} & \textbf{P5} & \textbf{P95} & $\boldsymbol{\rho(\text{RBB})}$ / Cohen's d \\
\midrule
\multicolumn{5}{l}{\textit{India} (n=1,957)} \\
Residual bubble breaking & 0.860 & -4.887 & 8.897 & -- \\
Residential deprivation & 21.40 & 12.86 & 37.00 & 0.199$^{***}$\\
Urban compactness$^a$ & 0.348 & 0.142 & 0.535 & 0.260$^{***}$ \\
POI entropy & 3.471 & 1.578 & 4.055 & -0.146$^{***}$\\
POI density & 0.001 & 0.000 & 0.004 & 0.010\\
Trip duration (min) & 69.58 & 50.27 & 103.2 & 0.032\\
Detour index & 1.695 & 1.018 & 3.739 & 0.026\\
\midrule
\multicolumn{5}{l}{\textit{Mexico} (n=2,601)} \\
Residual bubble breaking & 6.761 & -0.454 & 16.45 & -- \\
Residential deprivation & 27.04 & 10.62 & 43.51 & 0.155$^{***}$\\
Urban compactness$^a$ & 0.269 & 0.125 & 0.428 & 0.160$^{***}$ \\
POI entropy & 3.240 & 0.000 & 4.022 & -0.235$^{***}$ \\
POI density & 0.001 & 0.000 & 0.005 & -0.165$^{***}$\\
Trip duration (min) & 60.91 & 41.94 & 80.21 & 0.147$^{***}$ \\
Detour index & 1.373 & 1.244 & 1.706 & 0.141$^{***}$\\
\bottomrule
\end{tabular}\label{tab:stats}
\end{table}

\section{Additional modeling results}\label{sec:si_model}
\begin{figure}[!ht] %[H]
    \centering
    \includegraphics[width=0.7\linewidth]{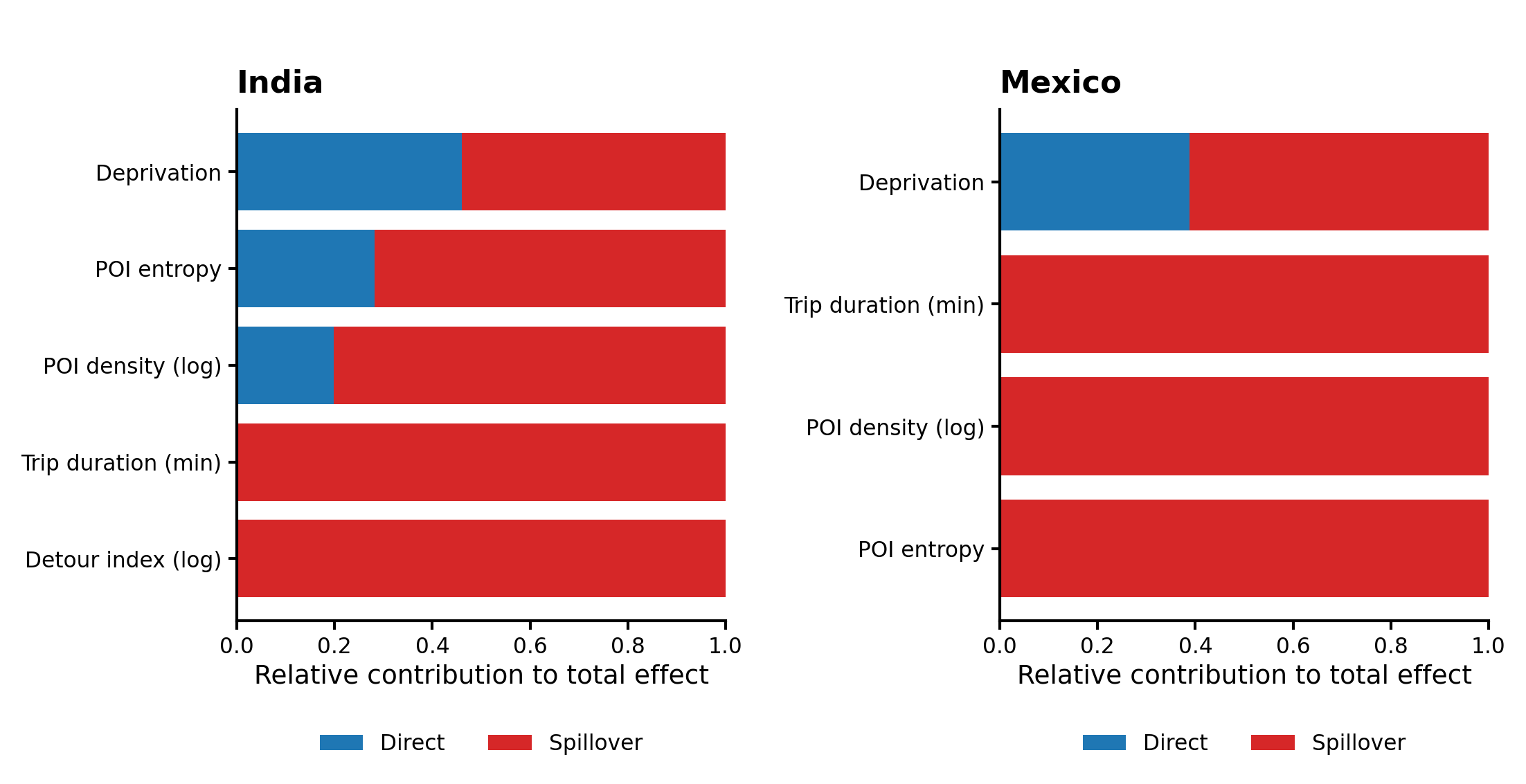}
    \caption{\textbf{Composition of SLX factor effects.} Relative contributions of direct and spillover effects to the total effect for each explanatory variable.}
    \label{fig:direct_vs_spillover}
\end{figure}

According to Supplementary Table \ref{tab:slx_comparison}, the SLX models reveal both local and neighboring contextual effects on residual bubble breaking in India and Mexico.
Supplementary Fig. \ref{fig:direct_vs_spillover} presents the decomposed direct and spillover effects of the significant coefficients from the SLX model for India and Mexico. \par

\begin{table}[ht]
\centering
\caption{
\textbf{SLX regression results for India and Mexico.} 
The dependent variable is the residual bubble-breaking (RBB). 
Models include local covariates and spatially lagged covariates constructed from first-order H3 contiguity. 
HC3-robust standard errors are reported in parentheses. $^{***}p<0.01$, $^{**}p<0.05$, $^{*}p<0.10$.
For India: $N=1957$, adjusted $R^{2}=0.347$, per-sample log-likelihood $=-2.68$, AIC $=5.38$, BIC $=5.41$. 
For Mexico: $N=2600$, adjusted $R^{2}=0.145$, per sample log-likelihood $=-3.02$, AIC $=6.04$, BIC $=6.07$.
}
\label{tab:slx_comparison}
\begin{tabular}{lrr}
\toprule
\textbf{Variable} & \textbf{India} & \textbf{Mexico} \\
\midrule
\textit{Direct} & & \\
POI entropy                  & $-$0.772$^{***}$ (0.231) & $-$0.050 (0.217) \\
POI density (log)        & 0.420$^{***}$ (0.155)    & $-$0.078 (0.174) \\
Deprivation  & 0.466$^{***}$ (0.023)    & 0.151$^{***}$ (0.018) \\
Detour index (log)        & 0.006 (0.056)            & $-$3.188 (2.933) \\
Trip duration (min)      & 0.006 (0.008)            & 0.006 (0.011) \\
Being in Compact city       & 0.676$^{***}$ (0.200)    & 1.307$^{***}$ (0.205) \\
\midrule
\textit{Spillover} & & \\
POI entropy               & $-$2.110$^{***}$ (0.346) & $-$3.122$^{***}$ (0.357) \\
POI density (log)      & 1.826$^{***}$ (0.253)    & 1.065$^{***}$ (0.265) \\
Deprivation     & $-$0.517$^{***}$ (0.028) & $-$0.231$^{***}$ (0.025) \\
Detour index (log)    & 0.337$^{***}$ (0.107)    & 4.887 (3.507) \\
Trip duration (min)   & 0.037$^{***}$ (0.013)    & 0.067$^{***}$ (0.019) \\
\bottomrule
\end{tabular}
\end{table}

\begin{figure}[!ht] %[H]
    \centering
    \includegraphics[width=1\linewidth]{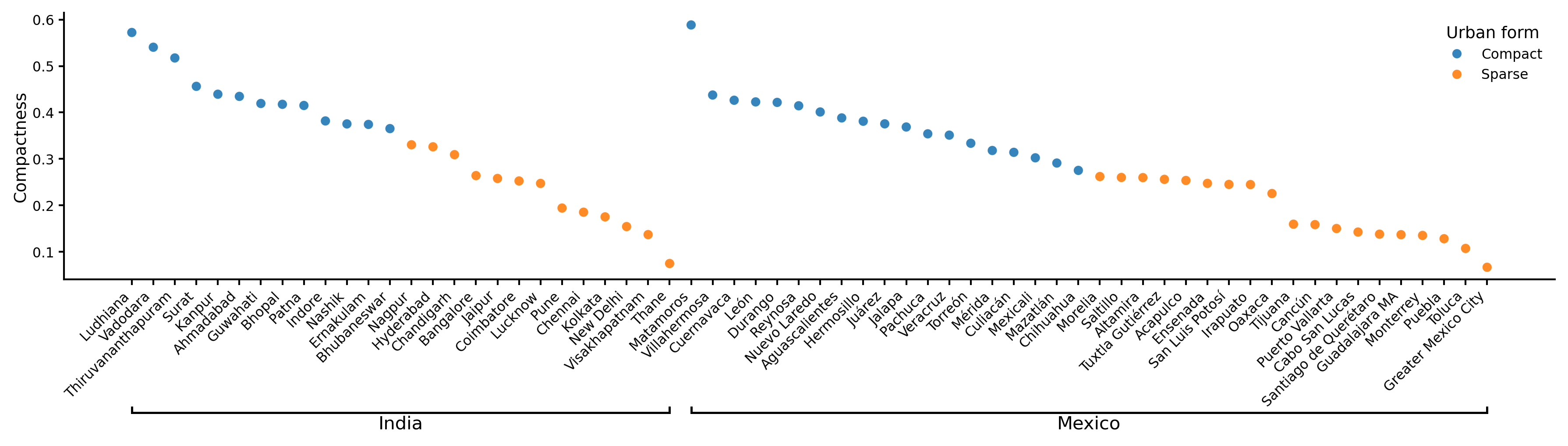}
    \caption{\textbf{Compactness of each city.} Sorted and colored by urban form group (Compact = blue, Sparse = orange).}
    \label{fig:urban_form}
\end{figure}

\begin{table}[ht]
\centering
\caption{
\textbf{Variance decomposition of residual bubble breaking explained by local and spillover factors.} 
The marginal contribution of each covariate group to the SLX model's explanatory power, measured as the reduction in $R^{2}$ when the corresponding local and spatially lagged terms are jointly removed from the full SLX specification. 
Larger $\Delta R^{2}$ values indicate greater importance for explaining variation in residual bubble breaking, conditional on all other variables.
}
\label{tab:slx_var_contributions}
\begin{tabular}{lrr}
\toprule
 & \multicolumn{2}{c}{$\Delta R^{2}$}\\
\textbf{Variable} & \textbf{India} & \textbf{Mexico} \\
\midrule
Deprivation  & 0.252    & 0.044 \\
POI entropy                  & 0.041 & 0.045 \\
POI density (log)        & 0.040   & 0.010 \\
Trip duration (min)      & 0.007            & 0.010 \\
Being in Compact city       & 0.004    & 0.014 \\
Detour index (log)        & 0.004            & 0.003 \\
\bottomrule
\end{tabular}
\end{table}

\section{Understanding POI density}\label{secb:density_puzzle}

To interpret the behavioral mechanisms underlying the positive association between POI density and residual bubble breaking identified in the SLX model, we conducted a series of stratified ordinary least squares (OLS) regressions with \emph{POI density} (log-transformed) as the dependent variable. 
This analysis aims to clarify what high POI density represents in terms of mobility volume, transport orientation, and local activity structure. \par

To identify regime-dependent meanings of POI density, areas were stratified by the median value of POI entropy within each country. 
This yields two subsamples per country: (i) \textbf{low-entropy areas} (below the national median), representing functionally limited environments, and (ii) \textbf{high-entropy areas} (above the median), representing functionally diverse environments. 
For each subsample, we estimated the model $\mathbf{y} = \mathbf{X}\boldsymbol{\beta} + \boldsymbol{\varepsilon}$, where $\mathbf{X}$ contains for each area $i$ the residential deprivation level, the number of generated trips per capita capturing mobility volume, and the number of transport POI per capita measuring transport infrastructure intensity. 
All models are estimated separately for India and Mexico using heteroskedasticity-robust (HC3) standard errors, with detailed results presented in Supplementary Table \ref{tab:ols_density_entropy_split}. \par

\begin{table}[ht]
\centering
\caption{
\textbf{Explaining POI density under low- and high-entropy regimes.} 
Separate regressions are estimated for India and Mexico under \emph{low entropy} (below median) and \emph{high entropy} (above median) subsamples. 
Robust (HC3) standard errors are reported in parentheses. 
$^{***}p<0.01$, $^{**}p<0.05$, $^{*}p<0.10$.
}
\label{tab:ols_density_entropy_split}
\begin{tabular}{llrr}
\toprule
\textbf{POI entropy regime} & \textbf{Variable} 
& \textbf{India} 
& \textbf{Mexico} \\
\midrule

\multirow{5}{*}{\textbf{High entropy}} 
& Intercept 
& $-10.070^{***}$ (0.265) 
& $-12.056^{***}$ (0.213) \\

& POI entropy 
& $0.759^{***}$ (0.069) 
& $1.444^{***}$ (0.059) \\

& Deprivation 
& $0.003^{*}$ (0.002) 
& $0.002$ (0.002) \\

& Trip number /capita
& $2.313^{***}$ (0.083) 
& $0.386^{***}$ (0.034) \\

& Transport POI number /capita
& $1949^{***}$ (216.4) 
& $1669^{***}$ (115.8) \\

\midrule

\multirow{5}{*}{\textbf{Low entropy}} 
& Intercept 
& $-10.593^{***}$ (0.118) 
& $-10.806^{***}$ (0.129) \\

& POI entropy 
& $0.836^{***}$ (0.038) 
& $1.026^{***}$ (0.036) \\

& Deprivation 
& $0.008^{***}$ (0.002) 
& $0.014^{***}$ (0.003) \\

& Trip number /capita
& $3.590^{***}$ (0.259) 
& $1.291^{***}$ (0.138) \\

& Transport POI number /capita
& $3809^{***}$ (563.6) 
& $1757^{***}$ (422.3) \\

\bottomrule
\end{tabular}
\end{table}

Transport-related POIs are manually extracted from Overture POI primary and secondary labels, including those such as bus stations, gas stations, car rental services, etc.

\section{Analysis on the impact of COVID-19 in India and Mexico}\label{secb:covid-19}

Major disruptions can alter how mobility translates into exposure, potentially amplifying existing differences across urban contexts. 
The COVID-19 pandemic provided a unique opportunity to examine this phenomenon and understand how external shocks affect urban mobility patterns and deprivation exposure dynamics \cite{yabe2020effects,yabe2023behavioral,cabrera_sustained_2025}. 
By analyzing mobility data from 15 June 2020 through December 2020, we investigated how pandemic restrictions influenced the breakdown of mobility networks across cities in India and Mexico -- two nations with divergent policy contexts, contrasting India's strict, nationwide lockdowns against Mexico's more flexible, regionalized measures \cite{mendez-lizarraga_evaluating_2022,padmakumar_covid-19_2022,perra_non-pharmaceutical_2021}.

\begin{figure}[!ht] %[H]
    \centering
    \includegraphics[width=1.0\linewidth]{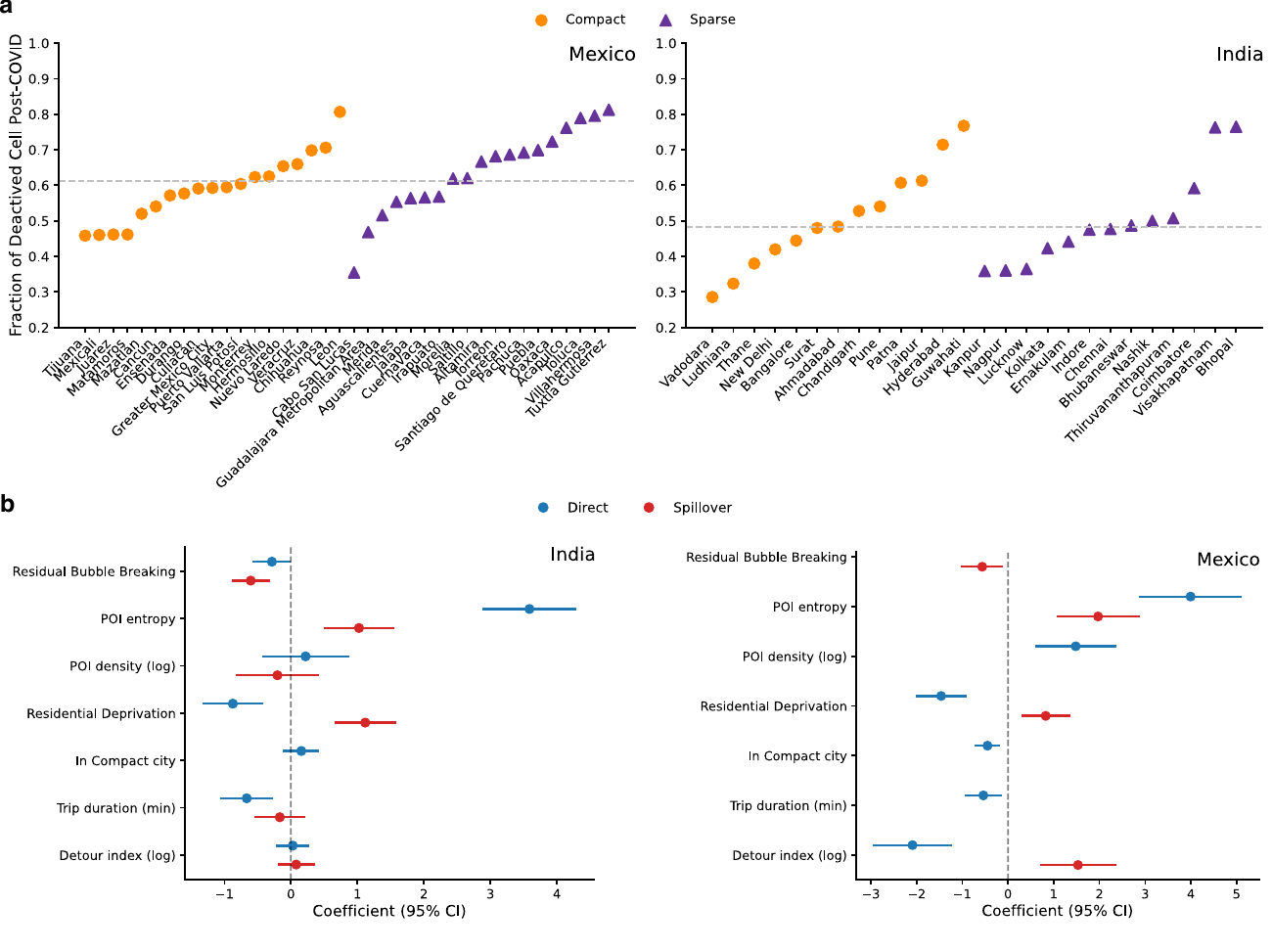}
    \caption{\textbf{COVID-19 mobility network fragmentation and predictors of area-level activity persistence across Mexican and Indian cities.} 
    \textbf{a}) Fraction of deactivated areas in each city across Mexico (left) and India (right), stratified by city compactness (orange compact, purple sparse). Horizontal dashed lines indicate the median fraction of deactivated areas within each country.
    \textbf{b}) Estimated normalized coefficients from stratified spatial logit models predicting area-level activity persistence post-COVID. Red bars represent spatial spillover effects, while blue bars show direct effects.
    }
    \label{fig_supp_disc}
\end{figure}

In the post-pandemic period, 57\% of previously active areas became disconnected (inactive). 
We define these disconnected areas as geographic areas that ceased regular movement connections with other parts of the city and no longer generated sufficient mobility outflows. 
This clear fragmentation of urban mobility networks indicates a shift toward highly localized and constrained activity spaces. 
We observed lower fractions of disconnected areas compared to sparse cities (See Figure \ref{fig_supp_disc} A). 
The emergence of such disconnection reflects multiple pandemic-related phenomena: stay-at-home orders, reduced economic activity, risk aversion to mobility, and the temporal concentration of essential trips \cite{perra_non-pharmaceutical_2021}.

To assess whether areas that depended on pre-pandemic mobility to overcome deprivation were more likely to deactivate under disruption, we estimated a stratified spatial logit model. 
The dependent variable equals 1 if the area remains present (active) in the post-COVID Origin-Destination matrix. 
We specified pre-pandemic residual bubble breaking as a key predictor, encoded as a binary indicator distinguishing areas above versus below the median. 
This allowed us to test whether the propensity to break bubbles independently predicts post-COVID activity, controlling for structural drivers such as residential deprivation, POI density and entropy, network detour index, and pre-pandemic travel times.

The model converged successfully for both countries (Mexico: pseudo-$R^{2} = 0.54$, $n = 2597$; India: pseudo-$R^{2} = 0.381$, $n = 1957$), explaining a substantial portion of variance in post-COVID activity patterns (See Figure \ref{fig_supp_disc} B).
The stratified models reveal a consistent separation between structural resilience and behavioral vulnerability.
Across both nations, POI entropy emerged as the strongest positive predictor of resilience (India: $\beta = 2.13$, $p < 0.001$; Mexico: $\beta = 1.77$, $p < 0.001$). 
This confirms that functionally diverse neighborhoods with urban forms supporting local living were significantly more likely to remain active during restrictions \cite{yabe2020effects,yabe2023behavioral}.
A high pre-pandemic residual bubble breaking significantly reduced the probability of post-COVID activity. 
However, this vulnerability exhibits distinct signatures across countries. 
In Mexico, the effect manifests exclusively through spatial spillovers ($\beta = -0.66$, $p = 0.01$), with no significant direct effect. 
Conversely, Indian cities show significant negative effects through both direct ($\beta = -0.29$, $p = 0.04$) and spillover pathways ($\beta = -1.01$, $p < 0.001$).

In Mexico, road network inefficiency -- captured via network detour index -- emerges as a critical secondary barrier to activity persistence ($\beta = -10.27$, $p < 0.001$), while it plays no significant role in India.
Ultimately, both countries displayed similar sensitivity to residential deprivation. 
The consistently stronger spillover coefficients for bubble breaking indicate that deactivation unfolded primarily at the neighborhood-cluster scale. 
This suggests that clusters of areas relying on extensive mobility to overcome deprivation were highly vulnerable to deactivation once movement was restricted, rather than vulnerability manifesting strictly through isolated area properties.
\end{appendices}
\end{document}